# A generative grammar of cooking

Ganesh Bagler*

Department of Computational Biology,
Infosys Center for Artificial Intelligence,
Indraprastha Institute of Information Technology Delhi (IIIT-Delhi),
Okhla Phase III, New Delhi 110020, India.
*Address for correspondence: bagler@iiitd.ac.in, ganesh.bagler@gmail.com

**Cooking is a uniquely human endeavor for transforming raw ingredients into delicious dishes**[1,2]**. Over centuries, cultures worldwide have evolved diverse cooking practices ingrained in their culinary traditions. Recipes, thus, are cultural capsules that capture culinary knowledge in elaborate cooking protocols**[3]**. While simple quantitative models have probed the patterns in recipe composition**[4–7] **and the process of cuisine evolution**[8–10]**, unlike other cultural quirks such as language**[11,12]**, the principles of cooking remain hitherto unexplored. The fundamental rules that drive the act of cooking, shaping recipe composition and cuisine architecture, are unclear. Here we present a generative grammar of cooking that captures the underlying culinary logic. By studying an extensive repository of structured recipes**[3]**, we identify core concepts and rules that together forge a combinatorial system for culinary synthesis. Building on the body of work done in the context of language**[11,12]**, the demonstration of a logically consistent generative framework offers profound insights into the act of cooking. Given the central role of food in nutrition and lifestyle disorders, culinary grammar provides leverage to improve public health through dietary interventions beyond applications for creative pursuits such as novel recipe generation**[13,14]**.**

Cooking is a creative act of turning nature into culture[2,3]. Starting with naturally available ingredients, the process of cooking transforms them into products that appeal to our sensory mechanisms and are a source of nourishment. This penchant for blending elements and processing them has led to the advent of the omnivorous human diet. Cultures around the world have evolved idiosyncrasies reflected in their culinary signatures—the use of unique ingredients and ways of processing and combining them. Notwithstanding the stylistic differences, the underlying process of cooking is quite similar across cultures. It is natural to compare cooking with language, another human quirk and cultural phenomenon with no equivalent in the animal kingdom. One exciting question that is hitherto unexplored is, "Is there a structure underlying the creative act of cooking?" One needs to investigate the corpus of prevailing cooking practices to answer this question.

**Documenting recipes.** Historically, civilizations have passed cooking instructions through generations by oral means. These would typically include an assortment of ingredients and a step-by-step protocol for processing. While recipes have been documented for a long time, the invention of printing in the 15th century provided the much-needed impetus for cataloging cooking practices. Cookbooks, however, became mainstream only a few centuries down the line. The credit for introducing the concept of standardized measuring, going beyond the qualitative description of cooking, is given to Fannie Farmer, the author of "The Boston Cooking-School Cook Book[15]." The advent of the internet democratized recipe compilation, stimulating the growth of online recipe sources. While human-readable these recipes aren't computable (Figure 1a). It is impossible to explore their underlying structure without the knowledge of different 'named entities' that appear in the recipe text. Natural language processing algorithms enable automated tagging of crucial culinary elements—ingredient name, quantity, unit, form (chopped, sliced, etc.), cooking action (roast, boil, sauté, etc.), and utensil, among others[16]. RecipeDB[3] is a structured compilation of over 118,000 recipes from cuisines across the globe (6 continents, 26 geo-cultural regions, and 74 countries), cooked using 268 processes (heat, cook, boil, simmer, bake, etc.) by blending over 23,500 ingredients from diverse categories. Such a repository, well-annotated for culinary terminologies with named entity recognition algorithms[16], captures contemporary cooking practices across cultures and provides a way to investigate fundamental cooking mechanisms.

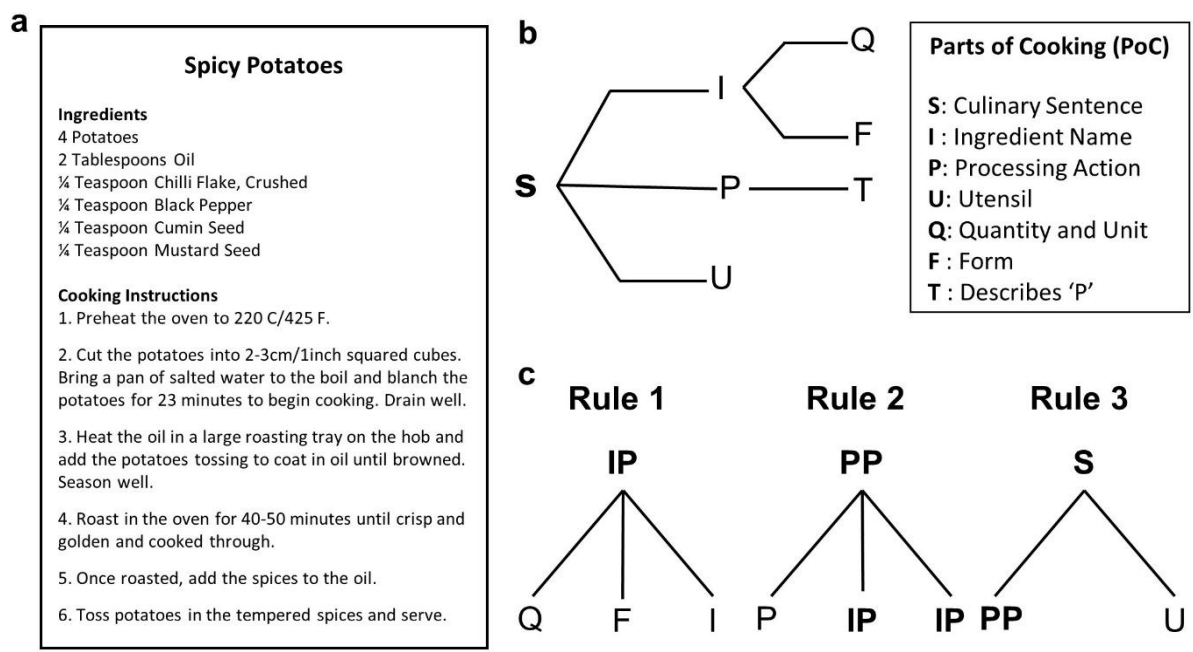

**Figure 1.** Cooking involves a rule-based blending of cubits, culinary concepts that capture essential aspects of cooking. (a) The structure of a modern recipe. A recipe description has three sections: title, ingredients section, and cooking instructions. (b) A cubit (culinary bit) represents culinary concepts found in the recipe description: culinary sentence (S), ingredient name (I), quantity-and-unit (Q), form (F), process (P), the qualifier of the process (T), and utensil (U). Analogous to parts of speech in a sentence, these could be called parts of cooking. (c) Three rules of culinary grammar. Rule 1, Rule 2, and Rule 3 define the ingredient phrase (IP), processing phrase (PP), and culinary sentence (S), respectively.

**Structure and evolution of cuisines.** In an effort to understand how cuisines evolve, some of the early computational studies[8–10] took a simplistic view of recipes. By modeling the recipe as an unordered list of ingredients devoid of nuances of cooking, they proposed a copy-mutate model reminiscent of genetic mechanisms. Other studies[4,5,7] included an additional layer of

detail—the flavor of ingredients[17] that serves as the basis for triggering olfactory and gustatory responses—to explore the hidden patterns in ingredient pairing. These studies revealed characteristic 'culinary fingerprints[6]' and a cuisine-specific bias in ingredient pairing rooted in their flavor profiles. While such data-driven investigations capture the plausible means by which recipe repertoire grows over time and the basis for the selection of ingredient combinations rooted in their flavor, they ignore crucial culinary factors—the quantity of ingredients, temporal flow, and processing actions—that play a major role in the preparation of a dish.

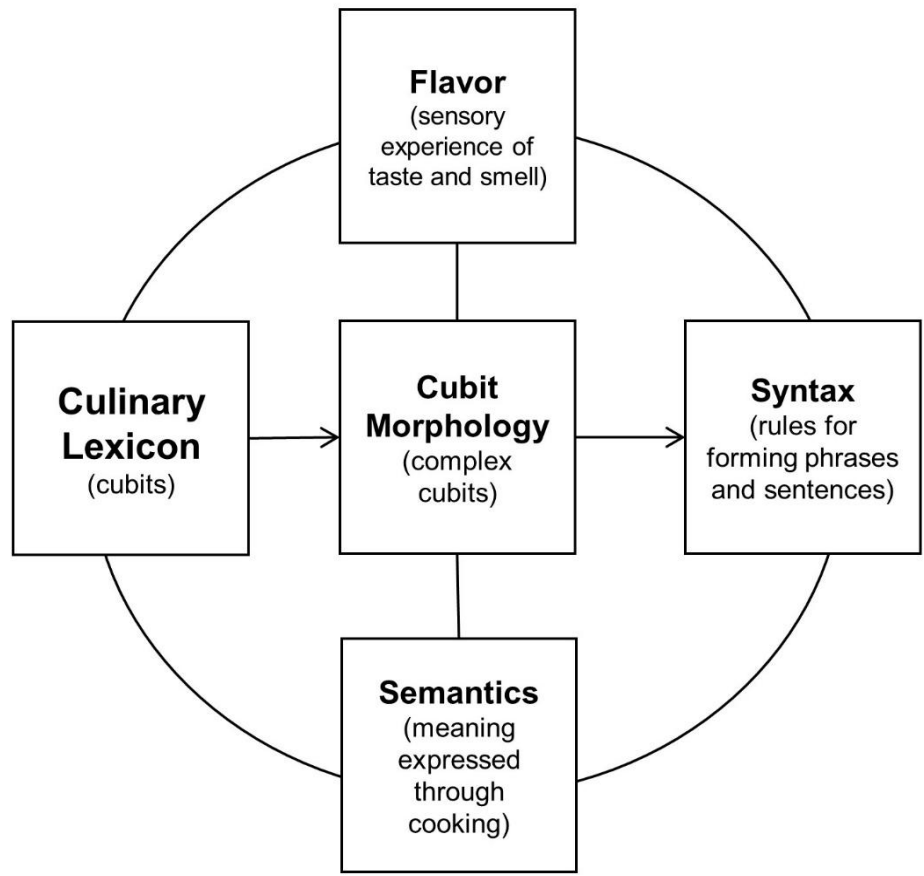

**Figure 2.** The anatomy of culinary grammar. The culinary grammar consists of the flavor, lexicon, morphology, syntax, and semantics of cooking.

**Cuisine as a language.** A story is to language what a recipe is to cuisine. Analogously, a word is for a sentence what an ingredient is for a statement of a recipe. The distribution of sentence sizes[18] and recipe sizes[4,5,7] share uncanny similarities. And the same is true for the popularity distribution of words[19] and ingredients[4,5,7]. Given that cuisines and languages share broad statistical features, it is tempting to drive this analogy further. A generative grammar framework of language[11,12,20] suggests that *words*, the basic building blocks of language, are combined following certain *rules* to yield sentences. The origin of this idea goes back to Panini[21], a fifth-century BC Sanskrit grammarian. Such grammar is a combinatorial system in which a small inventory of concepts can be assembled by rules into an immense set of distinct sentences[12]. Perhaps basic building blocks of cooking, similar to *words* and *rules* of a language[11,12,20], are 'culinary concepts' that are put together to create 'culinary sentences.' Subsequently, these 'sentences' coalesce to form richer structures—recipes. Let's name these 'culinary concepts' as 'cubits' (culinary bits)—the culinary equivalent of an information-theoretic bit (Figure 1b) (See Supplementary Information for a list of cubits). We suggest that analogous to *words* that relate to parts of speech, cubits are parts of cooking. We propose the following cubits: ingredient name (I), quantity-and-unit (Q), form (F), processing action (P), description of processing action (T), and utensil (U). While some of these are comparable to their linguistic counterparts (noun, determiner, adjective, verb), the others are not.

**Culinary grammar.** The grammar of cooking comprises a set of rules governing the flavor (a sensory experience), the lexicon of cubits, their morphology, syntax (patterns of arrangement of cubits), and semantics of cooking (Figure 2). Nature dictates the composition of an ingredient's flavor molecules, specifying its taste and smell. Further, led by the cultural practices, processing actions (P), in conjunction with cubits quantity (Q), form (F), and the way of processing (T), help alter the sensory experience of an ingredient. The culinary lexicon is a metaphorical kitchen shelf of cubits (I, P, U, Q, F, T) shaped by the forces of culture, climate, geography, religion, and history. The cubit morphology refers to the mechanisms that shape the cubits. Every ingredient has norms about its typical quantum in a culinary sentence other than its varieties and forms. The chilly is often 'cut' and used in much lesser quantity than a potato, for example.

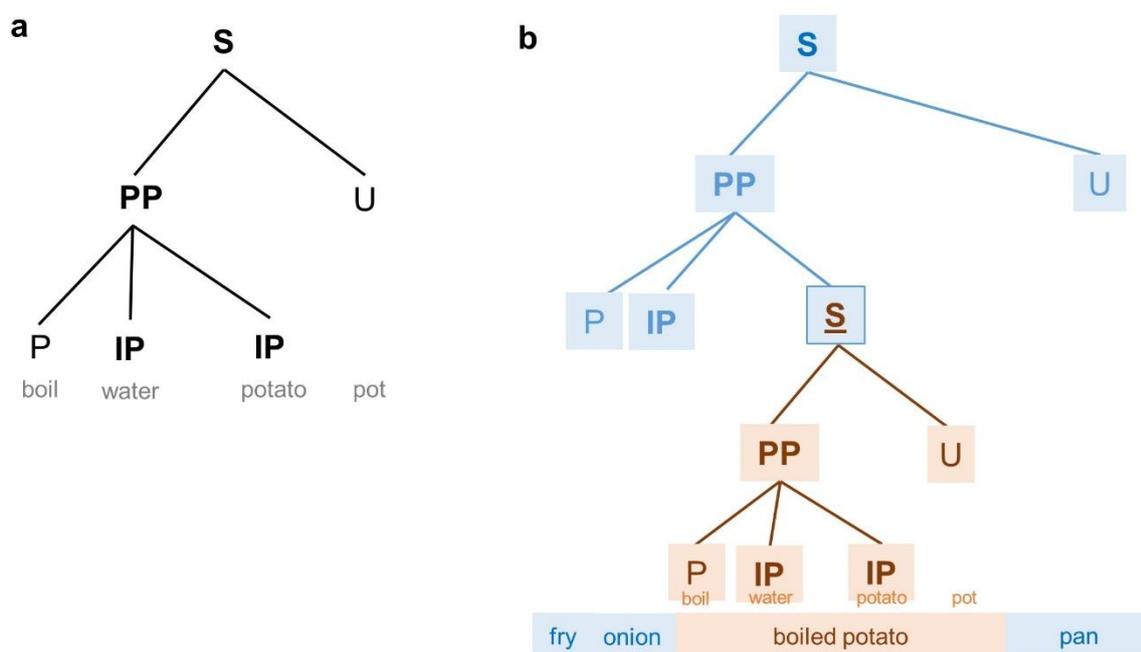

**Figure 3.** Culinary grammar illustrated in a simple cooking sentence. (a) A tree of cubits and phrases for a simple culinary sentence—Boil water [and] potato [in a] pot. (b) A tree with recursion, such that an ingredient phrase is a 'culinary sentence.' Here one of the ingredient phrases itself is a culinary sentence, thereby creating a recursive tree. The outer sentence ("Fry onion [and] boiled potato [in a] pan.") has an ingredient phrase (boiled potato) which appears as a standalone statement— "Boil water [and] potato [in a] pot."

The syntax presents a set of rules with which cubits combine to produce culinary phrases, which in turn are fused to yield a culinary sentence (Figure 1c). The first rule generates 'ingredient phrases' such as those listed in a recipe's 'ingredients' section (Figure 1a). An ingredient phrase (IP) may be composed of the cubits quantity-and-unit (Q), form (F), and the name of the ingredient (I). The second rule refers to the 'processing actions (P)' that produce chemical transformations in ingredients (I), profoundly changing flavor profiles. A processing phrase (PP) may consist of a process (P) cubit followed by its direct object(s), one or many ingredient phrases. Finally, the third rule yields gastronomically meaningful 'sentences.' A sentence in 'recipe instructions' may be composed of a processing phrase (PP) and a utensil (U) cubit. These rules, analogous to their linguistic parallels[12], are productive, abstract, and combinatorial. By assembling culinary concepts into phrases (according to the parts of cooking terms), this view of culinary structure gives an insight into the use and understanding of cooking. The proposed grammar materially simplifies cooking by introducing a 'culinary phrase structure' description.

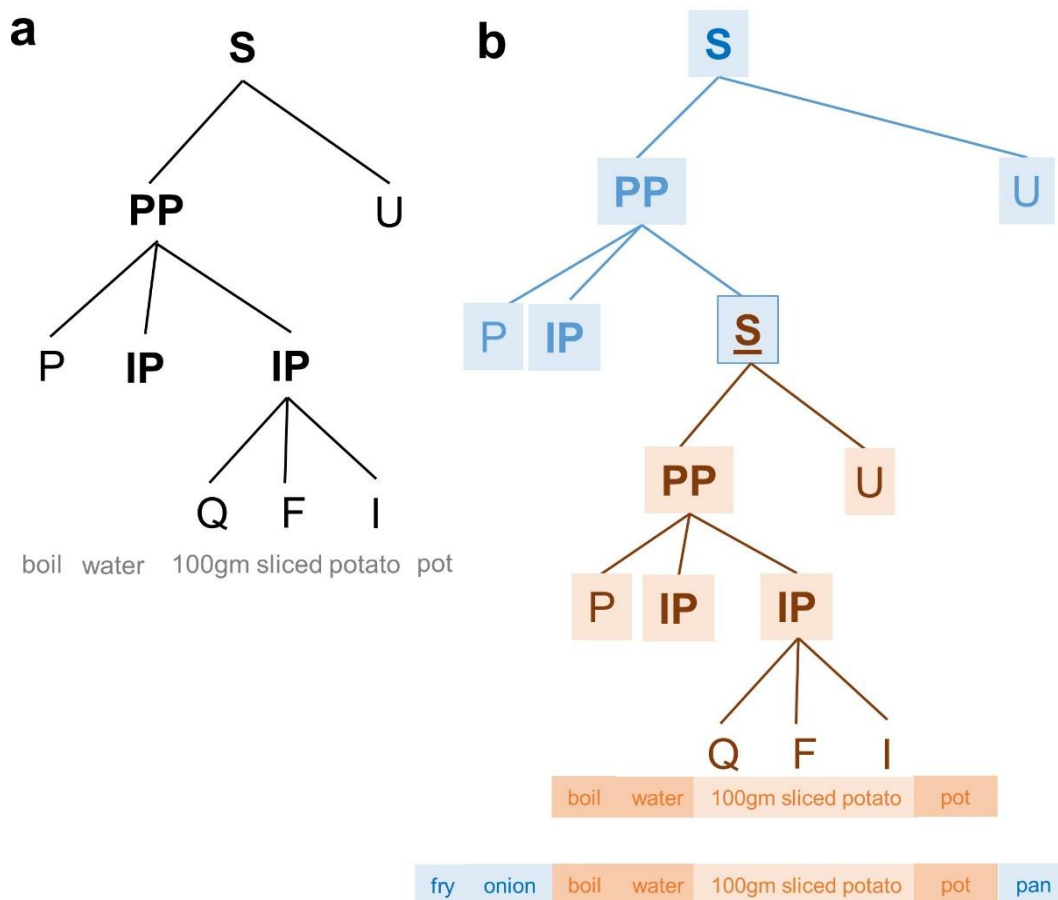

**Figure 4.** Culinary grammar illustrated to reflect nuances of cooking. (a) A tree of cubits and phrases for a culinary sentence invoking additional features of an ingredient phrase, namely, quantity-and-unit (Q) and form (F): "Boil water [and] 100gms [of] sliced potato [in a] pot." (b) A tree with recursion, such that an ingredient phrase is a 'culinary sentence.' Here a sentence is nested within a bigger sentence, "fry [the] onion [and] '100gm [of] boiled sliced potato' in a pan."

Accordingly, following the first rule, an ingredient phrase such as "100 gm of sliced potatoes" could yield a simple processing phrase such as "Boil the sliced potatoes," indicating a culinary transformation (Figure 3a). Finally, these culinary phrases could lead to a meaningful sentence—"Boil the sliced potatoes in a pot." We are aware that, in this simple form, the grammar assumes the presence of determiners (the, a) and prepositions (in). The beauty of such a structure lies in its ability to introduce the magic of recursion[12,22], which is at the heart of complex dynamical systems. Replacing an ingredient phrase with a sentence (Figure 3b) invokes this magic. Such recursive culinary grammar can express an unlimited number of distinct recipe steps, limited in practice only by stamina, money, and time. See Figure 4 for a nuanced version of the grammar to include details of ingredients (quantity-and-unit and form) and processing (descriptors such as thoroughly, slowly, for 10 minutes, etc.). Figure S1 (Supplementary Information) presents a nested tree depicting the process of preparing an "Italian Potato Salad."

**Combinatorial system for recipe generation.** Culinary grammar is a combinatorial system in which a small inventory of concepts can be assembled by rules into an immense set of distinct recipe sentences. Given the large variety in existing cubits (18,599 ingredient names, 71 quantity terms, 1299 forms, 270 processing actions, over 100 descriptors of processing actions, and 69 utensils), the number of ingredient and processing phrases and, subsequently, the number of such sentences is astronomical (See Supplementary Information for a list of cubits). Starting with such a kernel of gastronomic sentences, the number of generated recipes and hence the size of the cuisine can be unimaginably large. The beauty of such a combinatorial system is that it generates combinations that have never existed before in the culinary repository but that one might want to create someday. Thus, the creativity inherent in the act of cooking can be explained by a grammar of combinatorial rules through the infinite use of finite media[12].

> *Roasted meat,*
> *what a tasty feat.*
> *Cooking grows,*
> *by blending cubits.*

**The cooking machine.** By employing generative techniques, natural and cultural systems leverage the power of combinatorial mechanisms. The salient examples are DNA and protein sequences[23,24], language[11,12,20], and music[25,26]. A generative apparatus for creating variations, along with a selection layer for filtering out unfit variants reminiscent of Darwinian thought, seems to be the essence of dynamical, living systems. Having established a generative grammar for cooking, similar to the Turing's question[27], one of the most exciting thoughts is, "Can machines cook?" Deep-learning models can generate recipes with a decent novelty score[13,14] when trained on a large, annotated corpus of recipes (https://cosylab.iiitd.edu.in/ratatouille2). However, this approach relies on a statistical model devoid of insights into the anatomy of cooking. A generative cooking model would be of immense value for creating grammatical culinary phrases, sentences, and recipes, which can further be sifted for compliance with the culinary fingerprints[6]. Appropriate filtering strategies can help generate recipes that comply with dietary style, nutritional requirements, and inclusion or exclusions of certain ingredients to account for personal choices or the carbon footprint of recipes[28]. The generative framework of cooking has far-reaching consequences for 'computational gastronomy'[29] endeavors for achieving better public health and sustainability. Searching for palatable, tasty, nutritional, healthy, and sustainable recipes from a galaxy of 'grammatically generated recipes' has the power to transform the landscape of food. To quote Jean Anthelme Brillat-Savarin, the author of 'The Physiology of Taste'[30] and someone credited for coining the term 'gastronomy,' "The discovery of a new dish confers more happiness on humanity than the discovery of a new star."

**Acknowledgments.** The author thanks the Indraprastha Institute of Information Technology Delhi (IIIT-Delhi) for facilitating divergent cross-disciplinary thinking. Special thanks to the Department of Computational Biology and Infosys Center for Artificial Intelligence, which enabled the author to blend ideas from cooking, linguistics, biology, and information sciences to arrive at the generative framework. Putting together the computational gastronomy edifice (https://cosylab.iiitd.edu.in) would not have been possible for the author without prodigious efforts from his students, many of whom were undergraduates. Teaching the 'Computational Gastronomy' course to a computing-intense audience has been critical for the synthesis of key ideas. The author could not have reached this juncture without the stint at the Indian Institute of Technology Jodhpur (IIT Jodhpur), which was a tipping point in his intellectual journey.

**References**
1. Wrangham, R. *Catching Fire: How Cooking Made Us Human*. (Basic Books, 2010).
2. Pollan, M. *Cooked: A Natural History of Transformation*. (Penguin Books, 2014).
3. Batra, D. *et al.* RecipeDB: A resource for exploring recipes. *Database* **2020**, 1–10 (2020).
4. Ahn, Y.-Y., Ahnert, S. E., Bagrow, J. P. & Barabási, A.-L. Flavor network and the principles of food pairing. *Sci. Rep.* **1**, 196 (2011).
5. Jain, A., Rakhi, N. K. & Bagler, G. Spices form the basis of food pairing in Indian cuisine. *arXiv:1502.03815* 30 (2015).
6. Jain, A., Rakhi, N. K. & Bagler, G. Analysis of food pairing in regional cuisines of India. *PLoS One* **10**, (2015).
7. Singh, N. & Bagler, G. Data-driven investigations of culinary patterns in traditional recipes across the world. in *2018 IEEE 34th International Conference on Data Engineering Workshops (ICDEW)* 157–162 (2018).
8. Kinouchi, O., Diez-Garcia, R. W., Holanda, A. J., Zambianchi, P. & Roque, A. C. The non-equilibrium nature of culinary evolution. *New J. Phys.* (2008).
9. Jain, A. & Bagler, G. Culinary evolution models for Indian cuisines. *Physica A* **503**, 170–176 (2018).
10. Tuwani, R., Sahoo, N., Singh, N. & Bagler, G. Computational models for the evolution of world cuisines. in *35th IEEE International Conference on Data Engineering (ICDE 2019)* 85–90 (2019).
11. Chomsky, N. Three models for the description of language. *IRE Trans. Inf. Theory* **2**, 113–124 (1956).
12. Pinker, S. *Words and rules: The ingredients of language*. (Basic Books, 1999).
13. Agarwal, Y., Batra, D. & Bagler, G. Building Hierarchically Disentangled Language Models for Text Generation with Named Entities. in *28th International Conference on*

*Computational Linguistics (COLING)* 1–12 (2020).
14. Goel, M. *et al.* Ratatouille: A tool for Novel Recipe Generation. in *36th IEEE International Conference on Data Engineering (ICDE 2020), DECOR Workshop* 1–4 (2022).
15. Farmer, F. M. *The Boston Cooking-School Cook Book*. (Little, Brown and Company, 1901).
16. Diwan, N., Batra, D. & Bagler, G. A Named Entity Based Approach to Model Recipes. in *2020 IEEE 36th International Conference on Data Engineering Workshops (ICDEW)* 88–93 (2020).
17. Garg, N. *et al.* FlavorDB: a database of flavor molecules. *Nucleic Acids Res.* **46**, D1210–D1216 (2017).
18. Pilevar, M. T., Faili, H. & Pilevar, A. H. TEP: Tehran English-Persian parallel corpus. in *Lecture Notes in Computer Science (including subseries Lecture Notes in Artificial Intelligence and Lecture Notes in Bioinformatics)* 68–79 (2011).
19. Piantadosi, S. T. Zipf's word frequency law in natural language: A critical review and future directions. *Psychon. Bull. Rev.* **21**, 1112–1130 (2014).
20. Everaert, M. B. H., Huybregts, M. A. C., Chomsky, N., Berwick, R. C. & Bolhuis, J. J. Structures, Not Strings: Linguistics as Part of the Cognitive Sciences. *Trends in Cognitive Sciences* (2015).
21. Kadvany, J. Positional value and linguistic recursion. *J. Indian Philos.* **35**, 487–520 (2007).
22. Hofstadter, D. R. *Gödel, Escher, Bach: an Eternal Golden Braid*. (Penguin Books, 2000).
23. Searls, D. B. The linguistics of DNA. *Am. Sci.* **80**, 579–591 (1992).
24. Bralley, P. An introduction to molecular linguistics. *Bioscience* **46**, 146–153 (1996).
25. Lerdahl, F. & Ray, J. *A Generative Theory of Tonal Music*. (The MIT Press, 1996).
26. Baroni, M. Musical grammar and the study of cognitive processes of composition. *Music. Sci.* **3**, 3–21 (1999).
27. Turing, A. M. I.—Computer Machinery and Intelligence. *Mind* **LIX**, 433–460 (1950).
28. Piplani, P. *et al.* FoodPrint: Computing Carbon Footprint of Recipes. in *2022 IEEE 38th International Conference on Data Engineering Workshops (ICDEW)* 95–100 (2022).
29. Goel, M. & Bagler, G. Computational gastronomy: A data science approach to food. *J. Biosci.* **47**, 10 (2022).
30. Brillat-Savarin, J. A. *The Physiology of Taste*. (Vintage, 2011).

# SUPPLEMENTARY INFORMATION

**Figure S1** A nested tree of 'Italian Potato Salad' using the culinary grammar. (a) The recipe for 'Italian Potato Salad,' is annotated for its cubits. The ingredients section has seven ingredient phrases ($IP_1$, $IP_2$, $IP_3$, $IP_4$, $IP_5$, $IP_6$, and $IP_7$) and eight culinary sentences ($S_1$, $S_2$, $S_3$, $S_4$, $S_5$, $S_6$, $S_7$, and $S_8$). (b) Each of the seven ingredient phrases accounts for the details of quantity-and-unit (Q), form (F), and ingredient name (I). (c) A nested tree depicts the process of cooking the 'Italian Potato Salad' using cubits. Here, $S_8$ represents the complete recipe, which embodies multiple processing and ingredient phrases nested within. Some of these processing phrases are sentences themselves.

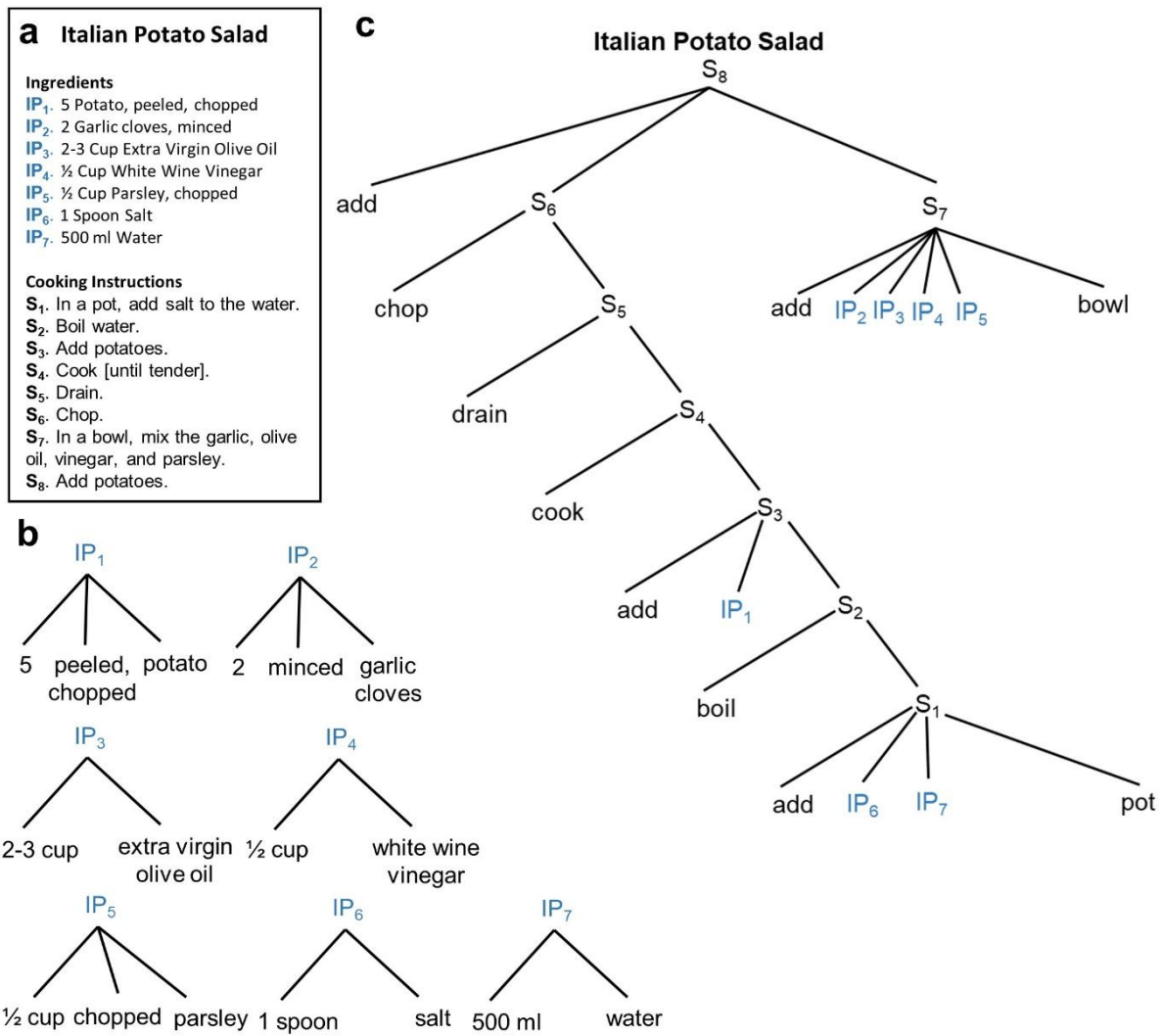

**Table S1.** Following is an illustrative list of 200 ingredients. The complete list of Ingredient name (I) cubits (18,599) that appear in RecipeDB is available **here**.

| | | | | |
|---|---|---|---|---|
| salt | onion | butter | water | garlic clove |
| olive oil | sugar | egg | tomato | black pepper |
| garlic | milk | pepper | salt pepper | cumin |
| vegetable oil | cinnamon | parsley | lemon juice | cilantro |
| ginger | carrot | soy sauce | flour | parmesan cheese |
| purpose flour | cream | oregano | beef | green onion |
| oil | potato | basil | brown sugar | lime juice |
| chicken broth | extra virgin olive oil | chicken breast | lemon | white sugar |
| chili powder | mushroom | garlic powder | paprika | bay leaf |
| red onion | cornstarch | cheddar cheese | thyme | celery |
| red pepper flake | baking powder | coriander | cayenne pepper | red bell pepper |
| tomato paste | vanilla extract | nutmeg | honey | white wine |
| salt black pepper | mozzarella cheese | chicken | tomato sauce | kosher salt |
| turmeric | green bell pepper | almond | sesame oil | clove |
| rice | red pepper | worcestershire sauce | heavy cream | shallot |
| cream cheese | scallion | bacon | mayonnaise | salsa |
| egg yolk | green chilies | chicken stock | zucchini | green pepper |
| spinach | coconut milk | lime | jalapeno pepper | shrimp |
| canola oil | black bean | sea salt | curry powder | cucumber |
| avocado | red wine | beef broth | yellow onion | orange juice |
| feta cheese | rosemary | flour tortilla | breadcrumb | vanilla |
| egg white | fish sauce | mint | raisin | black olive |
| cumin seed | baking soda | basil leaf | sesame seed | red wine vinegar |
| corn | coconut | cheese | lemon zest | walnut |
| pork | chicken breast half | white pepper | monterey jack cheese | dijon mustard |
| gingerroot | allspice | balsamic vinegar | vinegar | pea |
| italian seasoning | white onion | mustard | onion powder | peanut oil |
| corn tortilla | margarine | white vinegar | jalapeno | plain yogurt |
| ricotta cheese | eggplant | ketchup | flat leaf parsley | garam masala |
| rice vinegar | chive | peanut | dill | green bean |
| orange | caster sugar | mint leaf | cilantro leaf | bell pepper |
| pine nut | yeast | ham | plain flour | chicken thigh |
| cabbage | celery rib | banana | cardamom | lettuce |
| garlic salt | caper | leek | rom tomato | grain rice |
| swiss cheese | chickpea | apple | plum tomato | mango |
| cooking oil | white rice | italian sausage | cherry tomato | bread |
| yogurt | bean | chocolate | coriander leaf | cooking spray |
| lamb | spaghetti | strawberry | vegetable broth | whipping cream |
| cayenne | pecan | spring onion | black peppercorn | cold water |
| confectioner ' sugar | cocoa powder | kalamata olive | sherry | pineapple |

**Table S2.** Processing action (P) cubits (270) that appear in RecipeDB.

| add | heat | cook | stir | place | mix |
|---|---|---|---|---|---|
| cover | remove | serve | boil | simmer | bake |
| stirring | sprinkle | cool | preheat | cut | combine |
| drain | pour | set | season | beat | spread |
| transfer | whisk | smooth | top | blend | refrigerate |
| put | fry | melt | fold | reduce | dry |
| slice | coat | cream | chop | taste | peel |
| garnish | brush | press | drizzle | roll | grease |
| chill | saute | prepare | knead | stand | process |
| divide | toss | strain | puree | rinse | spray |
| seasoning | wash | arrange | store | sift | seal |
| rise | note | shape | wrap | soak | fill |
| roast | broil | stir-fry | dust | drop | dip |
| marinate | ladle | scoop | dice | mash | uncover |
| dissolve | separate | dressing | uncovered | pat | shake |
| turn | take | break | grind | toast | evaporated |
| steam | squeeze | crumble | scrape | shred | trim |
| scatter | reheat | flip | crush | dredge | clean |
| check | stuff | grate | whip | bubbling | punch |
| warm | skim | smoking | thread | wipe | decorate |
| marinade | zest | handle | rub | flatten | pull |
| push | reserve | mince | prick | open | pre-heat |
| butter | sear | measure | pressure | wok | split |
| rest | baste | bubble | burn | skin | pop |
| squash | move | soften | blanch | raise | deglaze |
| pierce | absorbed | stream | assemble | stack | ice |
| moisten | replace | pack | foil | tear | square |
| caramelized | smoke | throw | yield | save | pick |
| tie | invert | swirl | sieve | hold | poke |
| charred | floured | sizzle | wet | mound | dump |
| frost | test | distribute | thaw | wait | sit |
| glaze | scrub | carve | drip | stew | overcook |
| start | style | select | griddle | unroll | insert |
| sweat | tuck | meld | unwrap | scorching | caramelize |
| defrost | splutter | scramble | thicken | scald | twist |
| barbecue | unmold | poach | presentation | dress | smear |
| overmix | crockpot | smash | devein | moist | powdered |
| evaporate | flake | roux | slit | dripping | repeat |
| parboil | sprout | unfold | rising | stretch | steep |
| condensed | braise | whirl | minced | tilt | lard |

| deflate | strip | sort | dollop | submerge | rotate |
|---|---|---|---|---|---|
| splash | pan-fry | stop | curdle | char | freezing |
| slash | muddle | crimp | sterilize | blot | flavoring |
| foam | touch | snap | dash | immerse | settle |
| blitz | massage | sweeten | wilt | slather | mould |

**Table S3.** Utensil (U) cubits (69) that appear in RecipeDB.

| bowl | pan | oven | skillet | pot | saucepan |
|---|---|---|---|---|---|
| cup | dish | sheet | processor | fork | knife |
| plate | tablespoon | mixer | container | spatula | spoon |
| cooker | whisk | board | microwave | platter | sieve |
| boiler | jar | casserole | masher | skewer | frypan |
| cutter | strainer | stockpot | ladle | shaker | crockpot |
| peeler | crock | tray | blender | saucepot | basket |
| teaspoon | saucer | jug | ramekin | mug | bottle |
| kettle | beater | scoop | foil | paddle | blade |
| corer | processer | fryer | frying-pan | box | grinder |
| saucpan | sauceboat | shell | pestle | steamer | marzipan |
| disc | suacepan | basin | | | |

**Table S4.** The list of units from the Quantity-and-Units (Q) cubit (71) that appear in RecipeDB.

| cup | teaspoon | tablespoon | ounce | can | lb |
|---|---|---|---|---|---|
| package | clove | pinch | slice | gm | pound |
| bunch | dash | pods | jar | stalks | ml |
| quart | sprig | halved | piece | inch | pint |
| box | bag | bottle | loaf | packet | drop |
| envelope | kg | sheet | fluid ounce | leaves | head |
| cubes | carton | gallon | stick | scoop | bottles |
| thin | glass | container | square | ear | liters |
| dozen | scotch | cup halved | strip | can bottle | thick |
| cm | long | handful | round | splash | jar |
| thick slices | chips | big | slightly | thin slices | stems |
| cloves halved | pound halved | sprinkles | balls | few | |

**Table S5.** Following is an illustrative list of 200 forms. The complete list of Form (F) cubits (1299) that appear in RecipeDB is available **here**.

| chopped | ground | minced | sliced | diced |
|---|---|---|---|---|
| grated | cut | shredded | crushed | dried |
| sour | beaten | cooked | divided | drained |
| melted | softened | unsalted | crumbled | uncooked |
| cubed | sweet | peeled | peeled chopped | toasted |
| ground lean | granulated | peeled cut | peeled sliced | seeded chopped |
| skinless boneless | quartered | low fat | mixed | rinsed drained |
| canned | boiling | boneless | unsweetened | drained rinsed |
| low sodium | powdered | peeled diced | frozen | thawed |
| boneless skinless | seasoned | condensed | trimmed | sifted |
| smoked | separated | juiced | roasted | coarse |
| mashed | grated rind | peeled deveined | cracked | prepared |
| used | pitted | squeezed | ripe | evaporated |
| whipped | rinsed | seeded diced | refried | julienned |
| unbleached | chopped sweet | slivered | raw | washed |
| peeled cubed | boneless cut | seeded minced | flavored | boneless skinless cut |
| skim | mild | stewed | sweetened condensed | peeled minced |
| refrigerated | reduced sodium | diced drained | round | shredded blend |
| chilled | sliced drained | snipped | soft | trimmed cut |
| fat free | thawed frozen | peeled grated | smashed | boiled |
| diced undrained | rolled | pitted chopped | cooked drained | crushed dried |
| peeled quartered | warmed | shredded divided | seeded sliced | skinless |
| seedless | drained chopped | undrained | washed chopped | seeded cut |
| unsalted softened | unsalted melted | blended | bottled | distilled |
| semisweet | cooked chopped | boned | flaked | juice |
| peeled seeded chopped | blanched | fat | skinless boneless cut | seeded |
| coarse ground | semi-sweet | chopped divided | sun-dried | pressed |
| canned drained | halved sliced | salted | brewed | free |
| shelled | split | baked | chopped drained | firm |
| chopped cooked | cooked shredded | steamed | diced sweet | sliced sweet |
| peeled crushed | reduced fat | unpeeled | toasted chopped | pickled |
| unsalted cut | roasted chopped | unflavored | cooked diced | rind |
| chopped dried | lean | clarified | cooked crumbled | grated divided |
| trimmed sliced | packed | chopped toasted | sliced cut | fried |
| canned chopped | shaved | old | peeled cored sliced | cooked cut |
| creamed | unbaked | fat-free | shredded cooked | deseeded chopped |
| desiccated | peeled seeded diced | soaked | crumbled dried | cut round |
| part-skim | pitted sliced | sliced chopped | diced cooked | cut sweet |
| skinless cut boneless | canned diced | peeled cored chopped | corned | shelled deveined |
| sour low fat | chopped seeded | pureed | candied | cooked cubed |